\documentclass[12pt]{iopart}

\usepackage[colorlinks=true,citecolor=blue, linkcolor=blue, urlcolor=blue]{hyperref}

\usepackage{graphicx}
\usepackage{amssymb}
\usepackage{braket}
\usepackage[usenames, dvipsnames]{color}
\usepackage[caption=false,position=top,justification=raggedright,singlelinecheck=off]{subfig}

\newcommand \beq{\begin{eqnarray}}
\newcommand \eeq{\end{eqnarray}}
\def\Jvol<#1,#2,#3>{#1}
\def\Jpage<#1,#2,#3>{#2}
\def\Jyear<#1,#2,#3>{#3}

\begin{document}

\title[Nonequilibrium quantum dynamics of partial symmetry breaking]{Nonequilibrium quantum dynamics of partial symmetry breaking for ultracold bosons in an optical lattice ring trap}

\author{Xinxin Zhao$^{1,2}$, Marie A. McLain$^2$, J. Vijande$^3$, A. Ferrando$^4$, Lincoln D. Carr$^{2,5}$, and M.\'A. Garc\'{\i}a-March$^{2,6}$}
\address{$^1$International Center for Quantum Materials, School of Physics, Peking University, Beijing 100871, China    \\ $^2$Department of Physics, Colorado School of Mines, Golden, Colorado 80401, USA    \\ $^3$ Unidad Mixta de Investigaci\'{o}n en Radiof\'{\i}sica e Instrumentaci\'{o}n Nuclear en Medicina (IRIMED), Instituto de Investigaci\'{o}n Sanitaria La Fe (IIS-La Fe)-Universitat de Valencia (UV) Valencia and IFIC (CSIC−UV), Burjassot 46100, Spain    \\ $^4$ Department d'Optica. Universitat de Val\'{e}ncia, Dr. Moliner, 50, E-46100 Burjassot (Val\'{e}ncia), Spain    \\ $^5$ Physikalisches Institut, Universit\"{a}t Heidelberg, 69120 Heidelberg, Germany    \\ $^6$ ICFO - Institut de Ciencies Fotoniques, The Barcelona Institute of Science and Technology, 08860 Castelldefels (Barcelona), Spain}

%\date{\today}

\begin{abstract}
  A vortex in a Bose-Einstein condensate on a ring undergoes quantum dynamics in response to a quantum quench in terms of partial symmetry breaking from a uniform lattice to a biperiodic one.  Neither the current, a macroscopic measure, nor fidelity, a microscopic measure, exhibit critical behavior.  Instead, the symmetry memory succeeds in identifying the critical symmetry breaking at which the system begins to forget its initial symmetry state. We further identify a symmetry energy difference in the low lying excited states which trends with the symmetry memory.
\end{abstract}
%\pacs{}

\submitto{\NJP}

\maketitle

\section{Introduction}
\label{intro}

Nonequilibrium quantum dynamics is a rapidly growing field of study in part due to the emergence of hundreds of quantum simulator platforms build on multiple architectures, presenting enormous flexibility to explore new problems with detailed control of lattice structure, interaction strength, and bosonic or fermionic statistics~\cite{buluta2009,eisert2015quantum,molecular2016,memoryeffects}.  For example, global quantum quench dynamics have led to a deep understanding of the Kibble-Zurek mechanism relating non-equilibrium dynamics to critical exponents in quantum phase transitions~\cite{polkovnikov2011}.  Likewise, the study of local quenches or perturbations have taught us the role of short and long-range interactions in establishing a quantum speed limit on the propagation of correlations~\cite{cheneau2012light,richerme2014non}.  Use of a biperiodic optical lattice quenched to a uniform lattice has resulted in the first experimental demonstration of many-body localization~\cite{Schreiber2015,nandkishore2015many}.  Quantum simulators offer unusually isolated systems and long quantum coherence times, allowing careful exploration of the memory of initial conditions, and have resulted in e.g. approach to a new kind of ``thermal'' equilibrium under the eigenstate thermalization hypothesis, called the generalized Gibbs ensemble~\cite{Langen2015}.  By taking the opposite route from the many-body localization experiment, i.e., quenching from a uniform lattice to a bi-periodic one and thereby partially breaking the discrete rotational symmetry of a ring lattice, we find a completely different kind of long-lived robust dynamics in which newly identified quantum measures, the \emph{symmetry energy difference} and the \emph{symmetry memory}, reveal that the system only ``remembers'' its initial symmetry state below a critical partial symmetry breaking strength.

\begin{figure}
\centering
\includegraphics[width=0.85\textwidth]{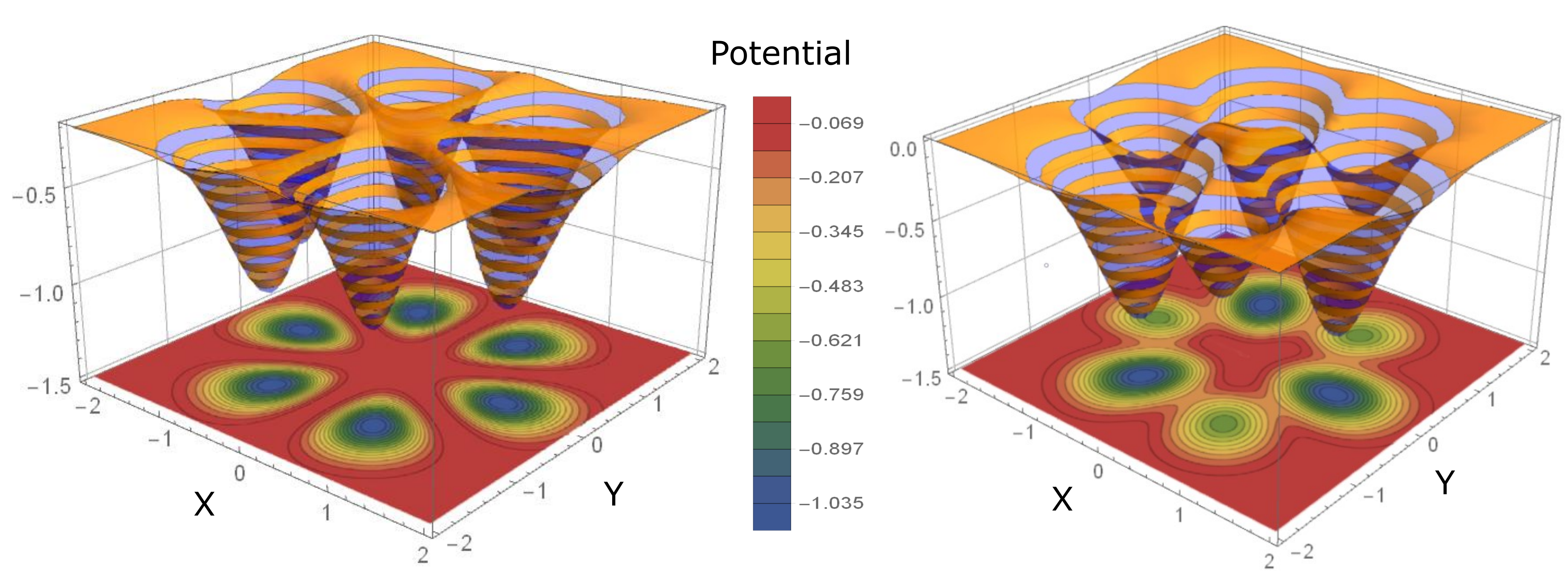}
\caption{\emph{Sketch of Partial Symmetry Breaking.} (left) Six-fold symmetric optical ring trap lattice potential. (right) Same potential after partial symmetry breaking, resulting in three fold symmetry.  The potential is shown partially transparent in orange and blue, and a projection appears on the 2D plane below in blue (low) to red (high) to help the reader visualize.  The ring may also occur as (left) the plaquette in a honeycomb lattice, as found e.g. in graphene, with (right) broken $A$-$B$ sublattice symmetry introducing a gap at the Dirac point. (Axes scaling are arbitrary).  We have sketched this particular case in arbitrary units.  Note that tunneling will be the same whether the valleys vary in height around the ring or the peaks, due to the form of the tunneling integral for overlap between quantum states localized on adjacent sites on the ring.  Both cases give rise to the same BHH.}
\label{fig:PSBtrap}
\end{figure}

Entanglement growth under a quantum quench is too rapid to capture long-time dynamics with tensor network methods~\cite{gobertD2005,kollathC2007}, as is necessary for large systems; therefore we employ exact diagonalization in small ring systems of $N=2$ to 10 bosons on $L=6$ to 10 sites\cite{Hilbertspacedim}, as well as perturbation theory to corroborate results and extrapolate trends in large interaction.  Partial symmetry breaking in the 6-site case in particular corresponds to breaking the $A$-$B$ sublattice symmetry in graphene, creating a gap at the Dirac point~\cite{geim2007,calandraM2007,Carrhoneycomb}.  We focus on the bosonic cold-atom-based quantum simulator architectures where much of the groundbreaking work on quantum dynamics has been performed and is frequently modeled with the Bose-Hubbard Hamiltonian (BHH)~\cite{bloch2012}.  In laser trapping of Bose-Einstein condensates, the discrete rotational symmetry ring trap~\cite{ringtrap,Campbell2018,Kalinikos2003,Beugnon2014} is an ideal potential to investigate dynamical symmetry breaking produced by a fast potential quench. Such a trap can be achieved by the interference of $XX$ and $YY$ Laguerre-Gaussian beams with the introduction of a quench to change the trap depth of even or odd sites thereafter, resulting in a functional form sketched in Fig.~\ref{fig:PSBtrap}.
In addition ``painted'' potentials with ultrafast lasers can achieve the same end~\cite{ryu2015,SQUIDpainted,Ryu2014}. Unlike the studies of the Berzinskii-Kosterlitz-Thouless (BKT) transition~\cite{fidelitymetric} and the fractional Mott insulator phase in the Bose-Hubbard superlattice~\cite{BHsuperlattice} which focused on the ground state (see also Supplemental material), it is necessary to go beyond fidelity, current, etc. to characterize the quantum dynamics of rotational states, or vortices in such potentials. We introduce the symmetry energy difference, a measure drawn from a cluster of low-lying excited states, and the symmetry memory, based on a time average over projections into rotational quantum numbers.  Such projections correspond to measurement of the winding number in the discretized ring system, exactly as occurs in BEC experiments. Although rotational measurements have typically been performed in the past on large continuous systems~\cite{stock2006}, with the advent of ultracold microscopy~\cite{bakr2010} and other precision techniques together with precise control over small systems~\cite{Langen2015}, rotational projections present an accessible avenue of exploration for upcoming quantum dynamics experiments. For example, our 6 site system may be taken as a study of the subsystem in a honeycomb lattice~\cite{soltanPanahi2011}, with an experiment performing an average over many such subsystems to do ``one-shot'' emulation of quantum averages.

\section{Partial symmetry breaking in a ring trap}
\label{psb}

In this section, we will gradually introduce partial symmetry breaking in an optical
lattice ring trap model, define the symmetry energy difference and the symmetry memory, and uncover a key dynamical critical behavior therein.

\subsection{Symmetry energy difference and symmetry memory}

The partial symmetry breaking Hamiltonian (PSBH), a rescaling of the usual BHH incorporating a two-period potential, takes the form
\begin{eqnarray}
\label{eq:PSBBHH}
\frac{\hat{H}_{\varepsilon}}{\bar{J}} = \frac{U}{2\bar{J}} \sum_{i=1}^L \hat{n}_i(\hat{n}_i-1)-\sum_{\langle i,j \rangle}(1+\varepsilon_{ij})(\hat{b}_{i}^\dagger\hat{b}_j+\hat{b}_{j}^\dagger\hat{b}_i)
\end{eqnarray}
where $U$ determines the on-site two-particle interaction; $\langle i,j \rangle$ denotes summation over the nearest neighbors; $\hat{b}_{i}^\dagger$ ($\hat{b}_i$) is the creation (annihilation) operator for bosons at site $i$ satisfying $[\hat{b}_i,\hat{b}_j^\dagger]=\delta_{ij}$; $\hat{n}_{i} \equiv \hat{b}_{i}^\dagger \hat{b}_{i}$ is the number operator; and $\varepsilon_{ij}\equiv\mathrm{Sign(i,j)}(J_{\rm e}-J_{\rm o})/(J_{\rm e}+J_{\rm o})$, with $|\varepsilon_{ij}|\in [0,1]$. The function $\mathrm{Sign(i,j)}\equiv\pm 1$ where the plus (minus) sign is taken for site $i$ even (odd). The hopping energy $J_{\mathrm{e}}$ ($J_{\mathrm{o}}$) encapsulates the biperiodic lattice through the usual overlap integral~\cite{lewensteinM2007}.  We scale our study to the average hopping energy $\bar{J}\equiv(J_{\rm e}+J_{\rm o})/2$, so that energies are in units of $\bar{J}$ and times in units of $\hbar/\bar{J}$.  Finally, we further define \emph{symmetry breaking strength} $\varepsilon=|\varepsilon_{ij}|$.  As we will show, there exists a critical $\varepsilon_c$ determining the vortex dynamics on the ring. The case of $\varepsilon=0$ restores the $L$-fold discrete rotationally symmetric lattice and the usual BHH whereas the introduction of $J_e,J_o$ enforces $L/2$-fold discrete rotational symmetry (we consider only even $L$ for simplicity). The hopping part of the PSBH model is similar to the Su-Schrieffer-Heeger (SSH) model under periodic boundary conditions, since the particles hop with staggered amplitudes in both models. However, they do have a significant difference: the SSH model has no on-site interactions, unlike the PSBH. It is interactions that generate entanglement between particles.  Moreover, we work in a range of interaction regimes including the strongly interacting one, where the dynamics are dominated by the interaction term.

\begin{figure}
\centering
 \includegraphics[width=0.78\textwidth]{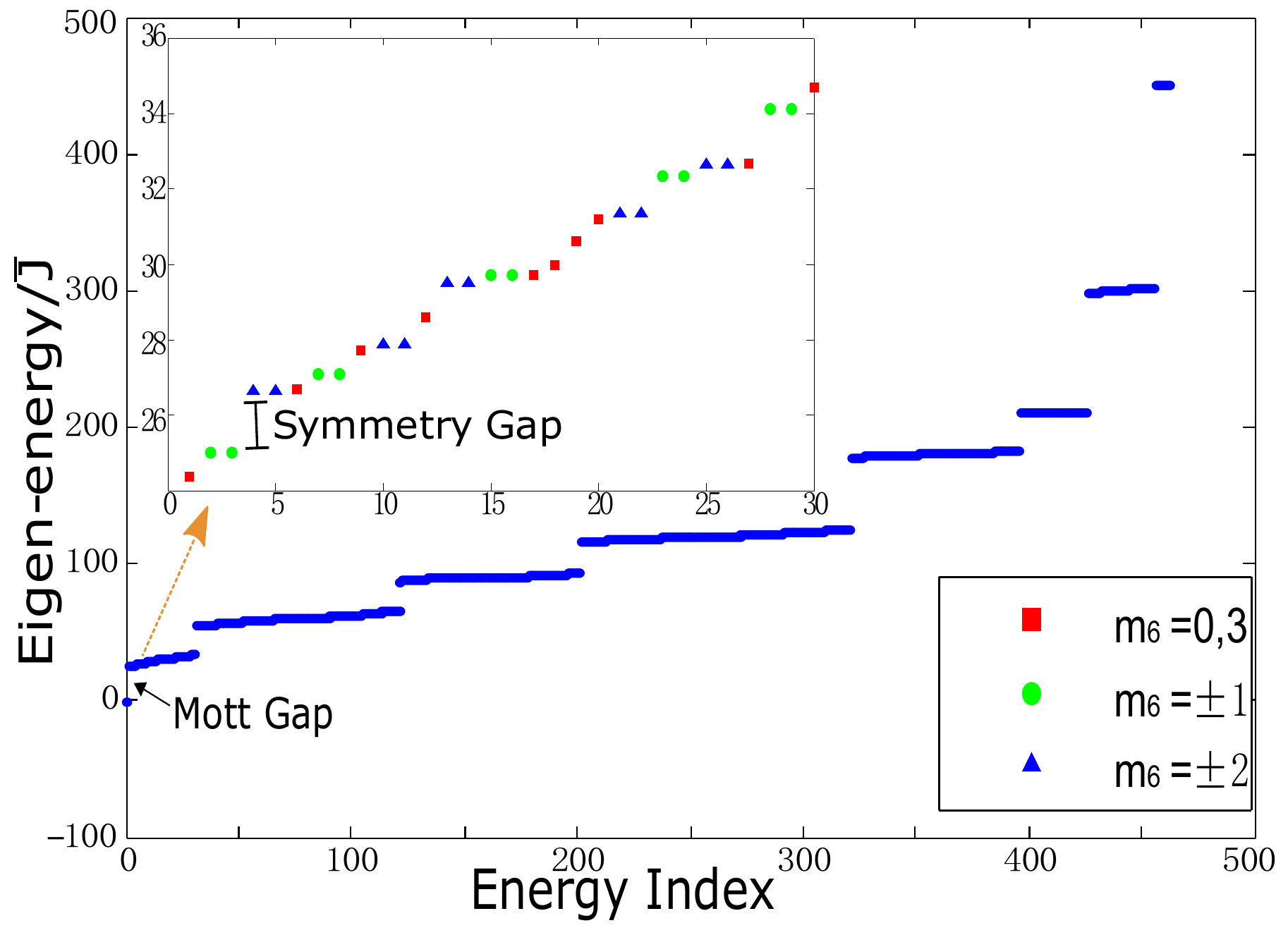}
 \caption{\emph{Energy Clustering and Symmetry Energy Difference.} The eigenenergy spectra for $N=6$ particles in a 6-fold rotationally symmetric ring trap with strong interactions, $U/\bar{J}=30$, separate into clusters lying above the Mott gap.  Insert: In the lowest lying cluster of excited states a new gap emerges, the \emph{symmetry energy difference}, a crucial measure of partial symmetry breaking which is key to characterizing vortex dynamics on the ring.}
\label{fig:sygap}
\end{figure}

Considering rotational eigenstates on the ring, the unitary $n$-fold discrete rotational symmetry operator satisfies $\hat{C}_n \ket{m_n}=e^{i 2 \pi m_{n}/n}\ket{m_n}$, where $m_{n}$ is the corresponding rotational quantum number, or winding number. Consider the honeycomb case $L=6$.  Then the quench procedure begins with $n=6$, $\varepsilon=0$ for $t<0$, and we take $n=3$, $\varepsilon \neq 0$, for $t\geq 0$. The PSBH has time reversal invariance symmetry.  Thus the eigenstates characterized by $\pm m_{n}$ are degenerate, as shown in Fig.~\ref{fig:sygap}, and the energy of the system depends only on $|m_{n}|$. For time $t>0$, i.e., after the quench, the $6$-fold symmetry is partially reduced to $3$-fold, and the energy eigenstates have a well-defined $m_{3}$ discrete rotational number. In group theory\cite{Ferrando2005}  the 3-fold rotational group $C_3$ is a subgroup of the 6-fold rotational group $C_6$, i.e., rotating a state under the $\hat{C}_6$ operator twice is equavalent to rotating it under the $\hat{C}_3$ operator once. Then for a rotational invariant state $\ket{i}$ embedded into a $C_6$ structure, one has $\hat{C}_3 \ket{i}=\hat{C}^2_6 \ket{i}=e^{i 2 \pi m_{6}*2/6}\ket{i}$. In general, for $\ell=2 n$, one finds $\hat{C}_\ell \ket{i}=\hat{C}^2_n \ket{i}=e^{i 2 \pi m_{n}*2/n}\ket{i}$. Considering an arbitary integer multiple of $2\pi$ phase on both sides, we obtain $e^{i 2 \pi m_{\ell}/\ell}=e^{i 2 \pi m_{n}*2/n}e^{i 2p \pi}$, where $p$ is an integer; in our 6-site case $m_6-m_3=3p$. Thus each $m_{3}$ corresponds to a pair of $m_{6}$ with distinct $p\pi$ phase differences under the action of the operator $\hat{C}_6$. For larger numbers of sites $n$, the same line of argument holds, and the symmetry of 8 sites can be partially broken to 4 sites, 10 to 5, and so on.

In the usual BHH, there are two well known distinct quantum phases, a Mott insulator and a superfluid; mesoscopic analogs of these phases exist in both canonical and grand canonical ensembles~\cite{Carr2010,sachdev1999}, where "quantum phase'' is determined by a sharp change in a quantum observable, rather than a singularity, as commonly observed in quantum simulator experiments~\cite{eisert2015quantum,cheneau2012light,richerme2014non,Schreiber2015}. A BKT transition occurs for integer filling around $(U/\bar{J})_{\rm crit}\simeq1/0.305\simeq3.28$~\cite{Ejima2011,Rigol2013} and a mean field $U(1)$ transition otherwise in the grand canonical ensemble. For our ring system of 6 to 10 sites, the mesoscopic analog of the critical point is between $(U/\bar{J})_{\rm crit}\simeq$ 5 to 10, depending on the choice of quantum measure used to determine the extremal behavior signifying the quantum phase transition~\cite{Carr2010}. We refer to regimes below (above) the effective critical point as weakly (strongly) interacting. The eigenenergy spectra of the BHH determined by exact diagonalization are shown in Fig.~\ref{fig:sygap}, for unit filling and the strongly-interacting case with a Mott gap, with $N=6$ particles on 6 sites. The eigenstates occur in clusters, in which several states are nearly degenerate; we refer to this as energy clustering. We can think of the partial symmetry breaking from 6-fold to 3-fold discrete rotational symmetry as mixing a defined $\hat{C}_6$ eigenstate $\ket{m_6}$ in the $L=6$ BHH with all states having the same value of $m_{3}$ under $\hat{C}_3$ acting on the 6 site ring, including compatible pairs of $m_{6}$. Thus, for example, under time evolution, for $\varepsilon\neq 0$ any $m_{3}=+1$ eigenstate of $\hat{C}_3$ will evolve as a linear combination of all the $m_{6}=+1,-2$ states, and the finite-size induced quantum recurrence is irrelevant to the time scales we study here. To capture this kind of evolution in partial symmetry breaking, it is expedient to trace its origin to the energy eigenspectrum in terms of the \emph{symmetry energy difference},
\begin{eqnarray}
\Delta_{s}=|E_{(m_{3}=+1,m_{6}=-2)}-E_{(m_{3}=+1,m_{6}=+1)}|
\label{eq:symgap}
\end{eqnarray}
which is the energy difference between the nearest compatible pair of $m_{6}$ states. A similar expression is obtained for larger system size.  Especially as system size grows, there are more kinds of symmetry energy difference in the spectrum: we find the lowest energy symmetry energy difference suffices to characterize the discrete rotational dynamics, similar in spirit to the use of Yrast states for the continuous rotational symmetry case~\cite{carr2010c}\footnote{We remind the reader that Yrast states are defined as the lowest energy state for fixed angular momentum, and play a key role in nuclear physics as well as ring BECs..}  We shall explain this point further in the time evolution process of symmetry memory and the relevant trends of symmetry energy difference and critical symmetry breaking strength in Sec.~\ref{ssec:critical}.  Note that the definition holds independent of filling factor and interaction regime. For small systems, e.g. for $N=6$, the symmetry energy difference takes the form of a gap between the two key symmetry states.  As shown in Fig.~\ref{fig:sygap}, a symmetry gap does indeed appear between the lowest $m_6=+1$ and $m_6=-2$ states in the eigenenergy spectra of an $N=6, L=6$ system.  Nevertheless, for larger systems like $N=12, L=12$, states with other rotational symmetry numbers may enter this gap between these still well-defined symmetry states.  We thus generically refer to $\Delta_{s}$ as a symmetry energy difference, rather than a symmetry gap. For example for $ N=12, L=12$, $\Delta_{s}(N=12, L=12)=|E_{(m_{6}=+1,m_{12}=-5)}-E_{(m_{6}=+1,m_{12}=+1)}|$, before finding the first $m_{12}=-5$ state above the lowest lying $m_{12}=+1$ state, a number of other $m_{12}=+1$ states may appear.  Similar issues can appear for Yrast states where different winding numbers may cross under rotational effects: however, the lowest lying states in each angular momentum manifold, and their energy difference, remain well defined.

 \begin{figure}
\centering
 \includegraphics[width=0.85\textwidth]{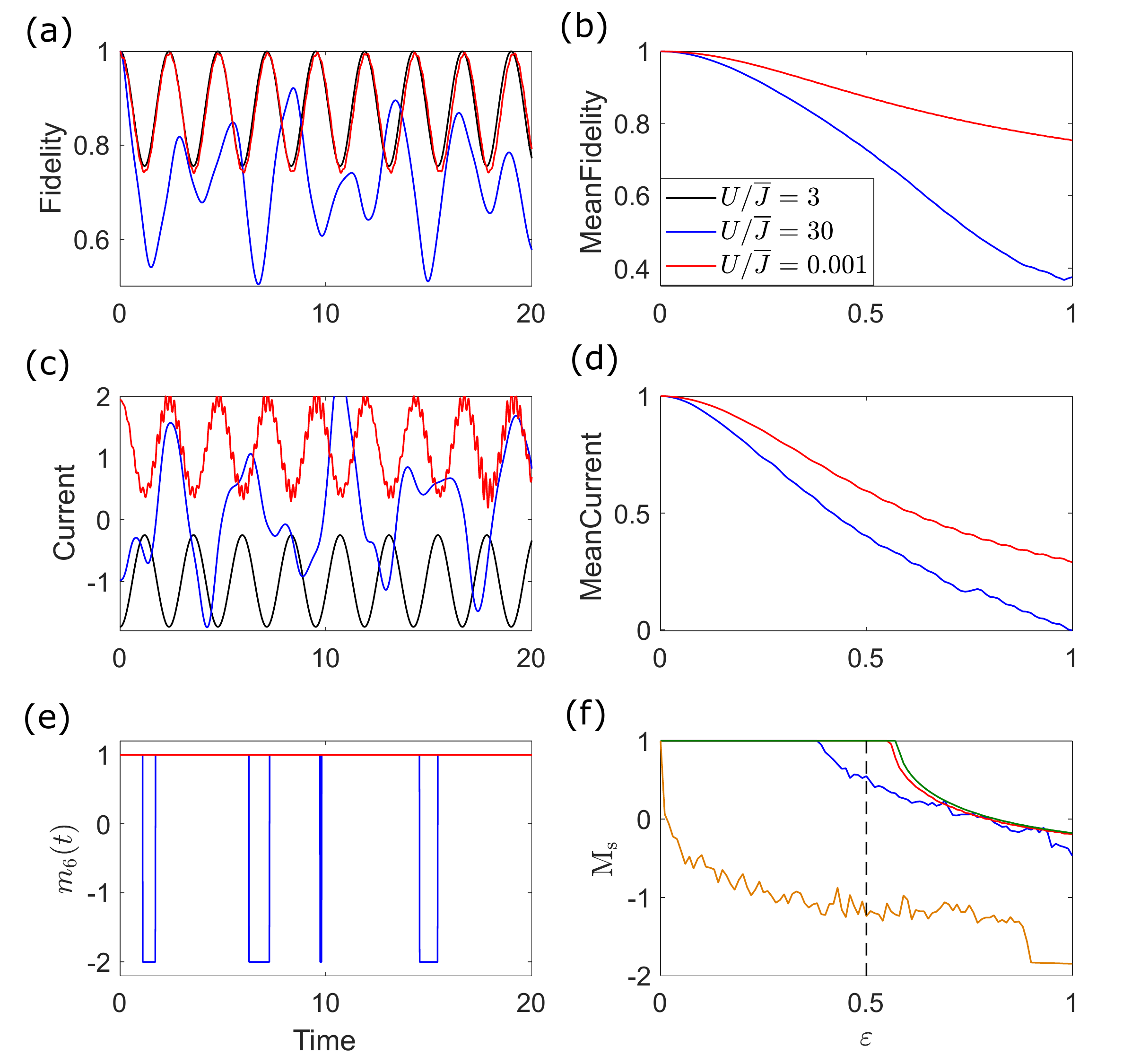}
 \caption{\emph{Microscopic, macroscopic, and symmetry-based quantum measures.}  Left: time evolution with interaction strength $U/\bar{J}=0.001, 3, 30$ (black, blue, and red curves). Right: time-average over 100 circuits around the ring vs. symmetry breaking strength. The (a)-(b) fidelity and (c)-(d) current do not exhibit critical behavior.  In contrast, (e) the rotational quantum number $m_6(t)$ measured by quantum projections constitutes a symmetry-based quantum measure and  (f) clearly exhibits critical behavior at $\varepsilon_c$ appearing as a cusp in the time average of the $C^6$ rotational symmetry operator, called the \emph{symmetry memory}, see Eq.~(\ref{eq:symemo}).  An attempt to reproduce critical behavior under mean-field time evolution is successful as expected for weak interactions (green curve), but utterly fails in the strongly-interacting case (orange curve). }
\label{fig:timeevo}
\end{figure}

We take our initial state to be a vortex of winding number +1: for $L=6$ this is $m_{6}=+1$, also the  lower excited energy state in the symmetry energy difference.  The time-dependent quantum average of the rotation operator is
\begin{eqnarray}
\label{eq:c6time}
 \bra{\psi(t)} \hat{C}_6 \ket{\psi(t)}=\eta (t) e^{i\frac{2\pi}{6} m_{6}(t)}
\end{eqnarray}
If $\ket{\psi(t)}$ is an eigenstate of $\hat{C}_6$, then $\eta(t)=1$ and $m_{6}(t)$ is a time-independent constant, which can only take values of $m_{6}=+1$ or $m_{6}=-2$ at each time point if we choose the lowest eigenenergy state $m_{6}=+1$ in the definition of the symmetry energy difference in our 6-site case as the initial state before real time evolution.  Once we quench to $\varepsilon\neq 0$,
$\ket{\psi(t)}$ retains its initial $m_{3}$ quantum number in terms of a superposition of $m_{6}=+1,-2$ states. At each time step $t$, the $m_{6}$ value of the six-fold rotation number $m_{6}(t)$ of $\ket{\psi(t)}$ is defined as the instantaneous projection of these two portions; $m_{6}(t)$ is consequently time-dependent.  Then, $\eta(t)$ becomes a real value between -1 and 1, which reflects the portion of these two six-fold rotational symmetry states. When $\eta(t)=0$, the $m_{6}=+1$ states and $m_{6}=-2$ states happen to occupy precisely equal portion in $\ket{\psi(t)}$. This brings out a neutral $\ket{\psi(t)}$, which does not show a six-fold rotational symmetry instaneously. However, numerically $\eta(t)$ will not be exactly zero. It is natural then to define the \emph{symmetry memory}

\begin{eqnarray}
M_{\rm s} \equiv \textstyle \frac{1}{\tau} \int_0^\tau dt \,m_6(t)\,
\label{eq:symemo}
\end{eqnarray}

In our test case of 6 sites, with the initial state is specified as $m_6=+1$, only states with symmetry properties $m_6=+1$ and $m_6=-2$ play a role in the time evolution of $|\psi(t)\rangle$.  Then the corresponding $m_6(t)$ jumps between $m_6=+1$ and $m_6=-2$, resulting in a symmetry memory valued between +1 and -2. As shown in Fig.~\ref{fig:timeevo}(f), critical behavior appears in this symmetry-based quantum measure, in strong contrast to the more typical microscopic measure, the time-averaged fidelity $\bar{f}/f(\varepsilon=0)\equiv\tau^{-1}\int_0^\tau dt |\bra{\psi(t)}\ket{\psi(0)}|$ in Fig.~\ref{fig:timeevo}(b), or the macroscopic measure, the time-averaged current $\bar{I}/I(\varepsilon=0)\equiv\sum_{j=1}^{L}(1+\varepsilon_{j,j+1})\tau^{-1}\int_0^\tau dt \bra{\psi(t)}i(\hat{b}_{j+1}^\dagger\hat{b}_j-\hat{b}_j^\dagger\hat{b}_{j+1})\ket{\psi(t)}$ in Fig.~\ref{fig:timeevo}(d) (where we've taken the lattice constant and $\hbar$ equal to unity.)  Although in Fig.~\ref{fig:timeevo}(e) the time evolution trend of $m_{6}(t)$ in the weakly-interacting case seems to be irregular in a few oscillation periods, the symmetry memory, as a statistical time average of $m_{6}(t)$, can be stabilized after a sufficiently long evolution time $\tau$. Therefore, we take the total simulation time $\tau$ as hundreds of the typical oscillation periods shown in Fig.~\ref{fig:timeevo}(a),(c),(e): this corresponds physically to hundreds of circuits of atoms around the ring. Typical hopping frequencies in BECs are kHz; thus $\tau$ is on the order of tens of milliseconds.

\subsection{Critical symmetry breaking strength}
\label{ssec:critical}

\begin{figure}[b]
\centering
\includegraphics[width=0.9\textwidth]{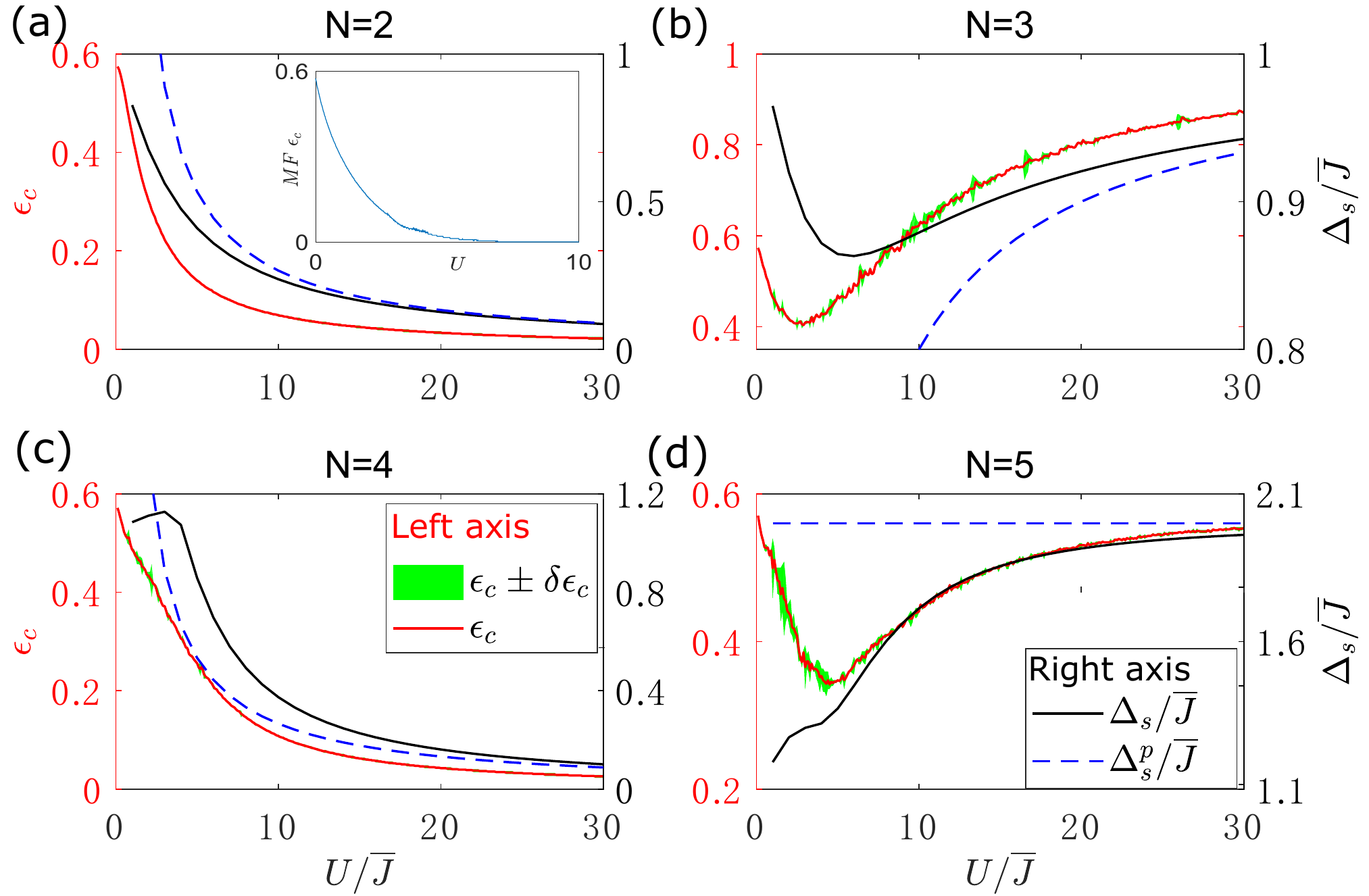}
\caption{ \emph{Trends in critical behavior with interaction strength.} Critical symmetry breaking strength $\varepsilon_c$ (left axis, red curves) trends with the symmetry energy difference $\Delta_s/\bar{J}$ (right axis, black curves) for strong interactions $U/\bar{J}$. Shown are $N=2,3,4,5$ particles on 6 sites. Perturbation theory (right axis, blue dashed curves) helps explain the two classes of asymptotic behavior in large $U/\bar{J}$ for different $N$, see text.  The green regions (left axis) attended with $\varepsilon_c$ indicate convergence error based on quadrupling total simulation time $\tau$ and higher resolution of $\varepsilon_c$. Insert in (a): Mean-field analysis showing complete quantum solution of the PSBH is necessary to obtain critical behavior beyond very weak interactions.}
\label{fig:critsygap}
\end{figure}

The time evolution of $m_{6}(t)$ in Fig.~\ref{fig:timeevo}(e) is distinct in the weakly and strongly-interacting cases at $\varepsilon = 0.5$. In the strongly-interacting case, $m_{6}(t)$ keeps its initial value of $m_{6}(t=0)=+1$ for all times and thus $M_{\rm s}=1$; however, in the weakly-interacting case, it occasionally loses the memory of its inital state and jumps to $m_{6}=-2$, since in this case only $m_6(t)=+1,-2$ are possible, generating $M_{\rm s}<1$. This difference exhibited in Fig.~\ref{fig:timeevo}(f) suggests that for each value of $U/\bar{J}$ a \emph{critical symmetry breaking strength} $\varepsilon_{\rm c}$ presents a cusp beyond which the symmetry memory $M_{\rm s}$ dips below unity, that is, beyond $\varepsilon_{\rm c}$ the system begins to lose the ability to retain its initial symmetry features. For mean-field analysis in Fig.\ref{fig:critsygap}(a, insert), only in the very weakly-interacting limit do we generate a similar critical behavior; the strongly-interacting case deviates severely from many-body predictions. Figure~\ref{fig:critsygap} shows that indeed critical behavior is exhibited for all interaction strengths.  What is the origin of this effect?

As shown in Fig.~\ref{fig:critsygap}, the symmetry energy difference from Fig.~\ref{fig:sygap} trends overall with $\varepsilon_c$.  This reveals the fact that the energy clusters seen in Fig.~\ref{fig:sygap} overall determine the trend in the dynamics, in particular the symmetry energy difference in the lowest cluster of excited states, as we here explain.  A detailed explanation appears in the time evolution of the symmetry memory, as we explain here for our test case of 6 sites. Recall that we take the lowest eigenenergy state in the symmetry energy difference $\Delta_s$ as the initial state for time evolution.  This lowest eigenenergy state is a simultaneous eigenstate of both the $\hat{C_6}$ and $\hat{C_3}$ rotational operators with rotational number $m_6=+1$ and $m_3=+1$. During real time evolution under the PSBH, the wavefunction $\ket{\psi(t)}$ can only contain states with rotational number $m_6=+1$ and $m_6=-2$, because these two kinds of state keep the same 3-fold rotational symmetry $m_3=+1$ in accordance with the system's unbroken 3-fold rotational symmetry.  Among these $m_6=+1$ and $m_6=-2$ states, the ones in the lowest cluster in the eigenenergy spectra in Fig.~\ref{fig:sygap} are easier to reach for $\ket{\psi(t)}$, since it takes the symmetry breaking less effort to conquer the energy difference between them and the initial state.  It follows that the lowest energy state with symmetry number $m_6=-2$ (the Yrast state for this winding number) has significant impact on changing the symmetry property of $\ket{\psi(t)}$. Thus the energy difference between the lowest energy two key symmetry states above the ground state, namely $m_6=+1$ and $m_6=-2$, captures much of the dynamics of vortices under a potential quench.  This is why the symmetry energy difference, Eq.~\ref{eq:symgap}, is so effective at predicting the critical partial symmetry breaking.

\subsection{Numerical methods and precision}
\label{ssec:precision}

In our calculations there are some practical issues affecting the numerics.  First, the symmetry energy difference can move up or down within the lowest energy cluster in Fig.~\ref{fig:sygap} for smaller interaction strengths, and this effect must be carefully accounted for, as we have done to obtain the curves in Fig.~\ref{fig:critsygap}.  This is because the crossing of different angular momentum manifolds is a function of interaction strengths.

There are also some numerical issues in the simulation. The total simulation time and the resolution of the symmetry breaking strength $\varepsilon$ affects the precision of the curves in Figs.~\ref{fig:timeevo},~\ref{fig:critsygap} and~\ref{fig:critlarge}. A sufficiently long total evolution time is required to ensure that transients pass and to ensure the time average is adequately sampling the jumps between symmetry states.  The optimal total evolution time varies with system size, but in general the larger the system size, the  bigger the Hilbert space, and thus the longer one needs to involve in time.  Nevertheless total times are reasonable on experimental time scales of at most a few hundred cycles around the ring. The exactness of $\varepsilon_c$  is directly linked to the total simulation time.  To estimate this effect, we quadruple the total simulation time $\tau$ and thus illustrate convergence error in Fig.~\ref{fig:critsygap} and Fig.~\ref{fig:critlarge}, shown as the green regions.

% \subsection{Truncation test}

\begin{figure}[b]
\centering
\includegraphics[width=0.93\textwidth]{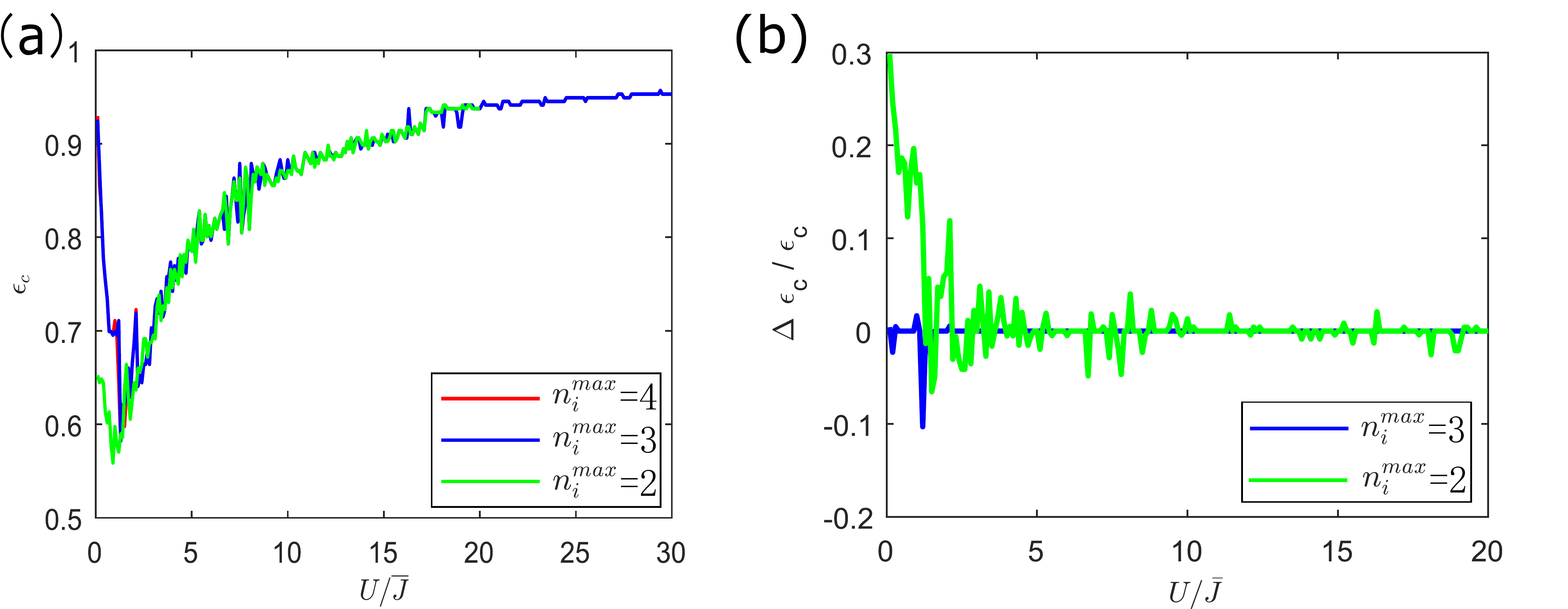}
\caption{ \emph{Truncation test of critical symmetry breaking strength.} (a) Critical symmetry breaking strength $\varepsilon_{\rm c}$ and (b) relative deviation of $\varepsilon_{\rm c}$ in a truncated system versus $\varepsilon_{\rm c}$ in its corresponding non-truncated system $\Delta \varepsilon_{\rm c}/ \varepsilon_{\rm c}$.  We study a range of interaction strengths $U/\bar{J}$, observing that for stronger interactions the error decreases.  The maximal error due to truncation effects is 20 to 30\% in the most extreme case of $n_i^{\mathrm{max}}=2$ particles allowed per site, and this greatly decreased by allowing up to 3 particles per site, with $n_i^{\mathrm{max}}=4$ the non-truncated case in our example of $L=8$, $N=4$.}
\label{fig:truncation}
\end{figure}

For the simulations in this Article, we restrict our study to exact diagonalization.  We clarify that not only is there a failure of approximation methods based on tensor networks at long times under quantum quenches~\cite{kollathC2007,cheneau2012light}, a well known problem making a strong case for nonequilibrium dynamics quantum simulator experiments, but also our potential quench in particular causes large fluctuation in on-site particle number and growth of entanglement. Thus the usual entanglement approximations applied in tensor network studies cause a failure in time evolution, in particular exploding norm of the reduced density matrix~\cite{schollwock2005,carr2012m}.  Although tensor network methods suffice for $\varepsilon \ll \epsilon_c$ where one expects an area law, in order to determine the critical point we must explore the region $\varepsilon \gtrsim \epsilon_c$ where a volume law holds, as for quantum quenches in general.  In this case tensor network methods fail.  Our avoidance of tensor network approaches is for a similar reason as many-body localization, where tensor network methods are not efficient to find the critical interaction strength for the transition from a volume law (thermalizing, smaller interaction side) to an area law (non-thermalizing, stronger interaction, many-body localized side).  Finally, exact diagonalization is certainly also necessary to obtain the complete energy eigenspectrum and determine the symmetry energy difference, as tensor networks are not efficient at obtaining trends in bands of excited states.

Thus restricting our studies to exact diagonalization to capture increasing entanglement, we instead explore another useful set of approximations, namely truncating the local on-site Hilbert space dimension.  This is necessary in order to simulate larger system sizes as otherwise the dimension of the Hilbert space grows as ${L+N-1}\choose{N}$.  We now take 4 particles on 8 sites shown in Fig.~\ref{fig:truncation} as a test case for a truncation test. In the strongly-interacting regime, the truncation works very well, with sub-percent error. In the worst case scenario for the $n_i^{\mathrm{max}}=2$ case in Fig.~\ref{fig:truncation}(b) and weak interactions, the error can be up to 20 to 30\%. This is because Fock states with up to $n_i^{\mathrm{max}}=N$ have significant weight in the dynamics. However, for our purposes $n_i^{\mathrm{max}}=2$ is sufficient to estimate general trends for larger system sizes.  We note that as number fluctuations are larger for smaller $L$, the case of $L=8$ presents a worst case scenario as we scale up to $L=18$.

\subsection{Perturbation theory}

\begin{figure}
\centering
\includegraphics[width=0.9\textwidth]{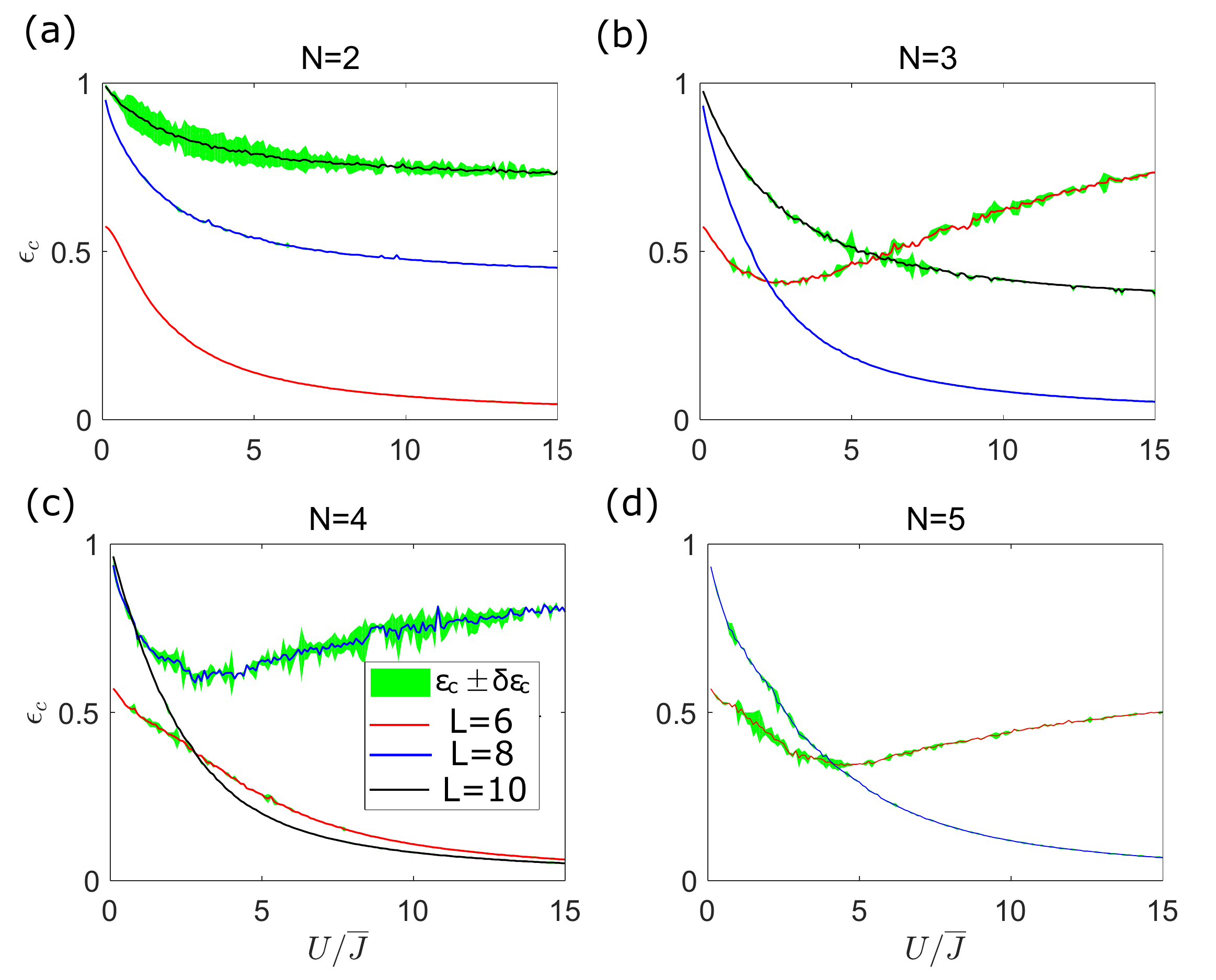}
\caption{\emph{Mesoscopic persistence of Critical Symmetry Breaking.} Critical symmetry breaking strength $\varepsilon_{\rm c}$ as a function of interaction $U/\bar{J}$ for $N=2,3,4,5$ particles on $L=6,8,10$ sites (red, blue, black curves).  Although exact diagonalization provides only limited access to larger systems, these results indicate critical behavior is pervasive and will be present even in the thermodynamic limit.  The same two classes of asymptotic trends for strong interactions are seen as in Fig.~\ref{fig:critsygap}.  The green error is determined in the same way as in Fig.~\ref{fig:critsygap}.}
\label{fig:critlarge}
\end{figure}

We observe for both $L=6$ and for larger systems in Fig.~\ref{fig:critlarge} two asymptotic trends for large interaction strength $U/\bar{J}$: $\varepsilon_c$ either (i) ascends to a non-zero constant or (ii) decreases toward zero, depending on $N$.  A brief study of second order degenerate perturbation theory on 6 sites reveals these two cases, taking the hopping term as a perturbation of the PSBH before the quench, and focusing on the symmetry energy difference $\Delta_s$ in Eq.~(\ref{eq:symgap}).  The symmetry energy difference $\Delta_s$ is generated before the quench, when the PSBH reduces to a standard BHH, $\hat{H} = \frac{U}{2} \sum_{i=1}^L \hat{n}_i(\hat{n}_i-1)-\bar{J} \sum_{\langle i,j \rangle}(\hat{b}_{i}^\dagger\hat{b}_j+\hat{b}_{j}^\dagger\hat{b}_i)$. Under perturbation in $\bar{J}$, $\Delta_{s}\simeq a(N)U+b(N)\bar{J}+c(N)\bar{J}^2/U$ to order $\bar{J}^2$.  $a(N)U$ is the zeroth order perturbation term, which comes from the BHH without a hopping term and depends merely on U. $b(N)\bar{J}$  is the first order perturbation term, which is determined by the hopping term according to the first order perturbation theory. $c(N)\bar{J}^2/U$ is the second order perturbation term, in which the energy is measured in unit of $\bar{J}^2/U$.  Rescaling to $\bar{J}$ to match the units used throughout this Article, $\Delta_{s}/\bar{J} \simeq a(N)U/\bar{J}+b(N)+c(N)\bar{J}/U$.

Because the upper and lower states in $\Delta_s$ are degenerate in the same energy cluster (see Fig.~\ref{fig:sygap}), the zeroth order term $a(N)=0$ for all $N$. Taking the 6-site system for example, we find $b(N)=0,1,0,2,1.58,4,0,2,0,4,2.97$ and  $c(N)=8/3,-2,8/3,0,3.15,0,8/3,-2,8/3,0,6.78$ for $N=2$ to 12, reducing more complicated expressions to numbers where necessary, to two decimal places. For $N=2,4,8,10$ we find $b(N)=0$. For these particle numbers the two states which determine the symmetry energy difference are still degenerate to first order in perturbation theory, and we must go to the second order coefficient $c(N)$. In contrast, for $N=3,5,6,7,9,11,12$, $b(N)\neq 0$, and for these particle numbers degeneracy is broken in the first order.  However, for completeness we calculate $c(N)$ for all $N$ of interest, where we find that, intriguingly, for $N=5,7,11$ the second order coefficient is zero.  Finally, $N=1$ is the trivial case, as there is no interaction in the problem, so perturbation theory in $U$ always yields zero.  We find that second order perturbation theory suffices to validate our simulations and obtain the trends in the strongly interacting limit. For case (i) in Fig.~\ref{fig:critsygap}(b) and (d), the nonzero first-order term $b(N)$ dominates $\Delta_s$, inducing an ascending trend. In contrast, $b(N)=0$ matches case (ii) in Fig.~\ref{fig:critsygap}(a) and (c), where the second-order term $c(N)$ is the major contribution and generates a decreasing trend.

We note that $b(N)=0$ or $c(N)=0$ are not due to an odd-even effect in particle number $N$. Instead, these cases are related to particular particle filling conditions, which decides how the eigenenergy states should be degenerate with each other.  In a 6-site system, the energy spectra of $N=9$ are similar to the $N=3$ case, and the energy spectra of  $N=2,4,8,10$ are also related (but not $N=6$, which is unit filling). The values of $b(N)$ and $c(N)$ in the perturbation analysis reflect this statement.

\subsection{Mean-field analysis}

To ascertain whether an analog of the critical symmetry breaking phenomenon discovered in the above sections exists in the mean-field theory, we explore mean-field simulations for the 6 site system, plotted in Fig.~\ref{fig:timeevo}(f) and Fig.~\ref{fig:critsygap}(a).  The mean-field approximation of the Bose-Hubbard Hamiltonian (BHH), i.e. discrete nonlinear Schrodinger (DNLS) equation in 1D lattice is

\begin{eqnarray}
\label{eq:MFBHH}
i \hbar \dot{\psi}_i=-J(\psi_{i+1}+\psi_{i-1})+gN|\psi_i|^2 \psi_i \,.
\end{eqnarray}
Here, compared to the corresponding standard BHH, the mean-field factor g*N corresponds to U in the BHH. The scalar wavefunction $\psi(t)$ is normalized to one. This DNLS can be obtained by propagating the field operator $\hat{b}_{i}$ forward in time using the BHH in the Heisenberg picture, meanwhile assuming the many-body state is a product of Glauber coherent states, $\langle \hat{b}_{i}^\dagger\hat{b}_i\hat{b}_i \rangle=\psi_i^*\psi_i\psi_i$, where $\psi_i\equiv \langle \hat{b}_i \rangle$. To describe a discretized ring lattice with 6 sites, we use polar coordinates to transfer the system to a 1D model. By rotating the uniform ground state of the DNLS, we can get states with typical rotational numbers, which can be used to detect rotational components in a certain state. The symmetry memory in the DNLS is defined in the same way as PSBH, see Eq.~(\ref{eq:symemo}). In Fig.~\ref{fig:timeevo}(f), we obtain mean-field time evolution for both weakly and strongly-interacting cases, shown in green and orange curves respectively. It is clear that for the weakly-interacting limit the mean-field reproduces a similar critical behavior, however, for the strongly-interacting case, the mean-field results strongly deviate from the many-body Bose-Hubbard model. Thus our results, while limiting nicely to a mean-field effect, are not at all inherently mean-field, but extend throughout the strongly-interacting regime, and do not depend on the assumption of Glauber coherent states or a single dominant single-particle mode with an accompanying phase.

\subsection{Partial symmetry breaking is distinct from the fractional- and unit-filling Mott insulator to superfluid transition}

\begin{figure}[t]
\centering
\includegraphics[width=0.95\textwidth]{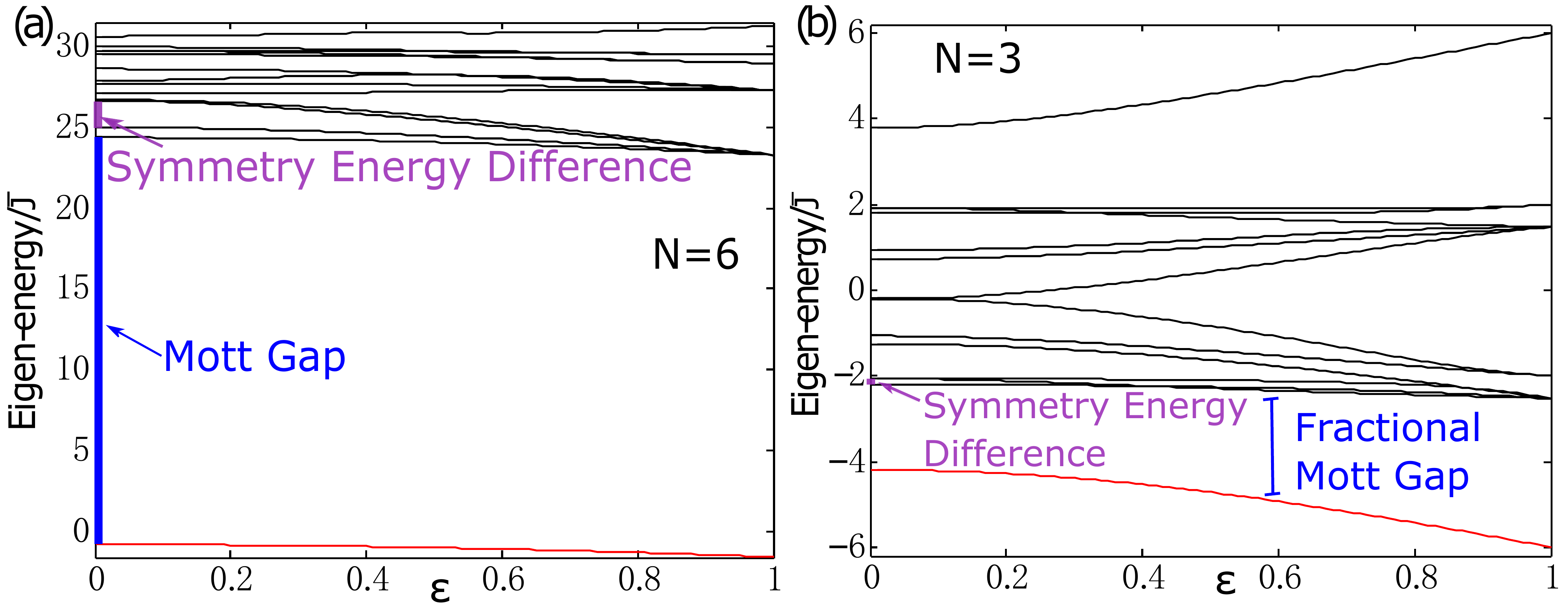}
\caption{\emph{Eigenenergy spectra of partial symmetry breaking Hamiltonian as a function of the symmetry breaking strength $\epsilon$}. The two plots are for ground states (red curves) and all the excited states (black curves) in the first energy cluster in 6-site systems in the strongly-interacting case, here $U/\bar{J}=30$.  In both plots, the symmetry energy difference and the Mott gap is difined at $\epsilon=0$. We track their evolution when $\epsilon$ varies from 0 to 1.  (a) For unit filling of total particle number $N=6$ in the strongly-interacting limit, the symmetry energy difference (purple bar on vertical axis) is far beyond the Mott gap (blue bar on vertical axis). (b) For half filling of total particle number $N=3$ at strongly-interacting limit, the symmetry energy difference is always beyond the possible fractional Mott gap.  Thus the symmetry energy difference, occurring in excited states, is an entirely distinct phenomenon from the Mott gap associated with the ground state.}
\label{fig:mott}
\end{figure}

In the Bose-Hubbard Hamiltonian in the strongly-interacting case, the Mott phase and a Mott gap emerge for integer filling as shown in Fig.~\ref{fig:mott}(a). After partial symmetry breaking with a hopping modulation, the system becomes a Bose-Hubbard superlattice. Further study of the ground state phase diagram ~\cite{BHsuperlattice} showed that fractional-filling Mott insulator phases could appear in the superlattice; in this partial symmetry breaking case, half-filling Mott insulator phases can likewise occur. In our study we have introduced a \emph{symmetry energy difference} $\Delta_s$ to describe the critical symmetry breaking phenomenon and used the lower eigenenergy state in the symmetry energy difference as the initial state, which is an excited state in the eigenenergy spectra, see Fig.~\ref{fig:sygap} in the article. In contrast, the superfluid phase, Mott phase and the fractional-filling Mott phase are related to the system's ground state, and bear no relation to gaps appearing in the first cluster of excited states. The symmetry energy difference and both the integer and fractional Mott gaps are located at completely different positions in the energy spectra, as indicated in Fig.~\ref{fig:mott} with blue and purple bars, respectively. Since the system size in Fig.~\ref{fig:mott} is small, the symmetry energy difference turns into a symmetry gap therein.

In our study, the wavefunction is evolved under the partial symmetry breaking Hamiltonian (PSBH). Thereafter, for a 6-site lattice after the quench, the wavefunction becomes the combination of all the states with the same $\hat{C}_3$ rotational number as the initial state. Since the ground state has a different $\hat{C}_3$, it will never participate in the evolution, once again showing neither the integer nor fractional Mott gaps play a role in the dynamics. Figure~\ref{fig:mott} shows the energy difference between the ground state and the lowest excited state which has the same $\hat{C}_3$ rotational number as the initial state. Therefore, the energy level of the symmetry-based phenomenon is always beyond the ground state, i.e., it won't be controlled or affected by the Mott insulator-superfluid transition. In conclusion, the symmetry breaking in our study bears no relation to the emergence of the fractional-filling Mott phases, and this symmetry breaking is distinct from the Mott insulator-superfluid transition.

\subsection{Finite size scaling}

\begin{figure}[b]
\centering
\includegraphics[width=0.95\textwidth]{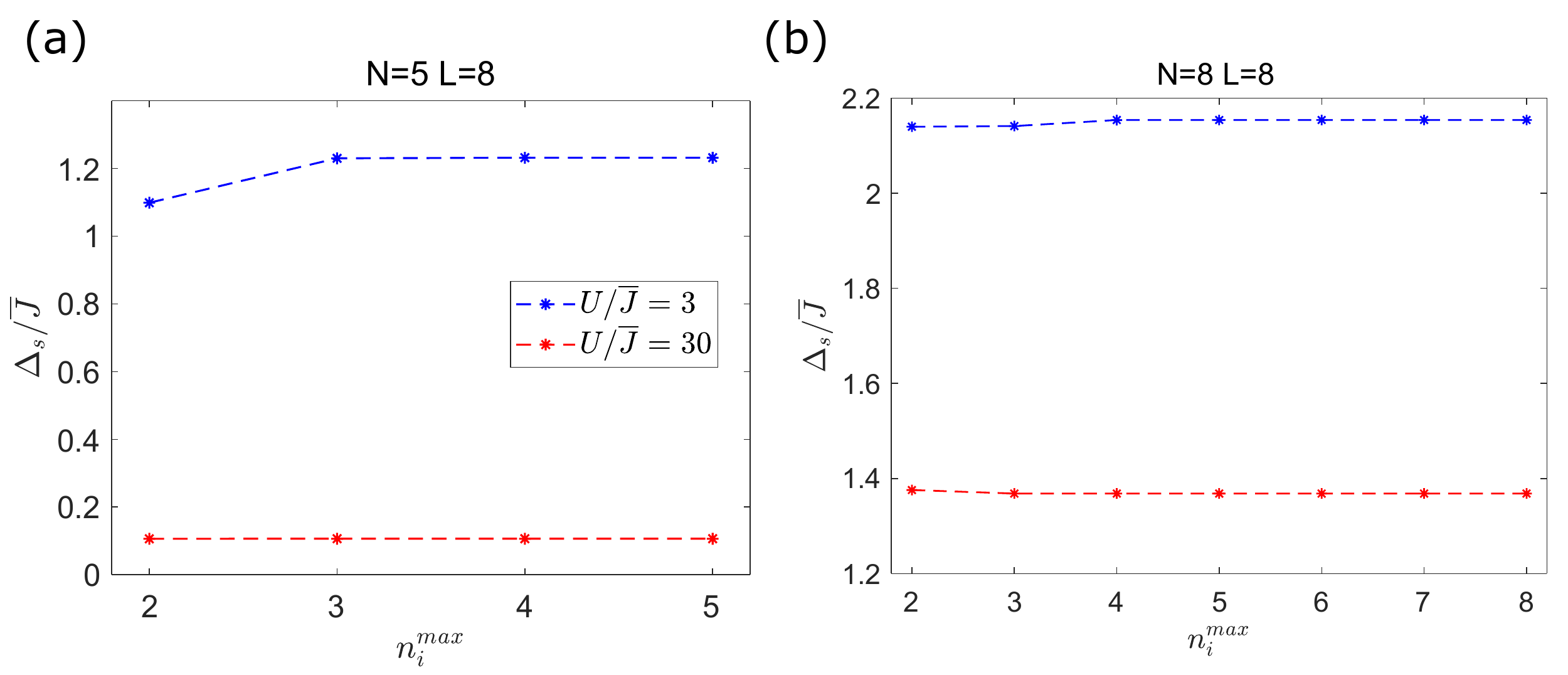}
\caption{\emph{Truncation test of symmetry energy difference.} Symmetry energy difference $\Delta_s/\bar{J}$ under truncation of the on-site Hilbert space with (a) $n_i^{\mathrm{max}}=2,3,4,5$ for $N=5, L=8$ and  (b) $n_i^{\mathrm{max}}=2,3,4,5,6,7,8$ for $N=8, L=8$. In the strongly-interacting regime (red curves), $\Delta_s$ is nearly constant with $n_i^{\mathrm{max}}$. In the weakly-interacting regime (blue curves), $\Delta_s$ exhibits visible but small variations with smaller $n_i^{\mathrm{max}}$.}
\label{fig:truncatesg}
\end{figure}

\begin{figure}[t]
\centering
\includegraphics[width=0.95\textwidth]{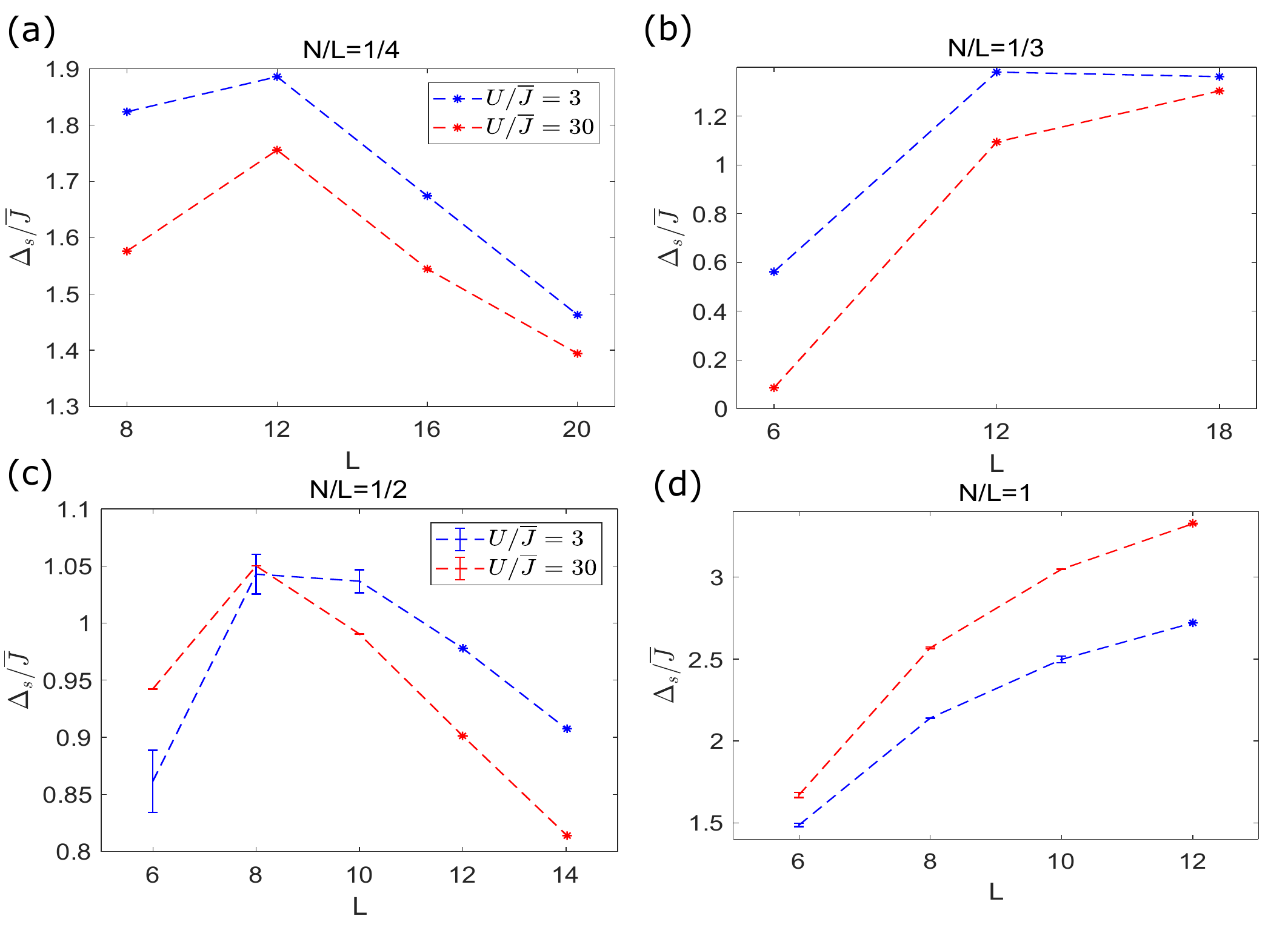}
\caption{\emph{Finite size scaling analysis of symmetry energy difference.} The density N/M is fixed at (a) 1/4, (b) 1/3, (c) 1/2, (d) 1 in each panel with increasing lattice size L. The maximum occupation number per site $n_i^{\mathrm{max}}=2$ is truncated, see Fig.~\ref{fig:truncatesg}. Red curves represent the strongly-interacting cases $U/\bar{J}=30$, and blue curves represent the weakly-interacting cases $U/\bar{J}=3$. Error bars for L=6, 8, 10 in (c) half filling and (d) unit filling for both strongly and weakly-interacting cases indicate truncation error is overall small.  The symmetry energy difference remains large for systems of up to 20 sites, indicating at minimum a persistent mesoscopic effect.}
\label{fig:scaling}
\end{figure}

Thus far our analysis has focused mainly on the $L=6$ case study.  Although restricted to exact diagonalization, as discussed in Sec.~\ref{ssec:precision}, we can make use of truncation of the local Hilbert space to push somewhat into larger systems.  Truncation is performed via restricting the maximum occupation number per site, $n_i^{\mathrm{max}}$. For the purposes of illustration, our truncation tests are shown for two cases in Fig.~\ref{fig:truncatesg}, $N=5, L=8$ and $N=8, L=8$. Both cases demonstrate that in the strongly-interacting regime the truncation effect is negligible, since the symmetry energy difference $\Delta_s$ is nearly constant with decreasing $n_i^{\mathrm{max}}$ (red curves). However, similar to the conclusion in Fig.~\ref{fig:truncation} for the weakly-interacting regime, $\Delta_s$ varies obviously in smaller $n_i^{\mathrm{max}}$ cases (blue curves). Compared with the largest $n_i^{\mathrm{max}}$ cases, these variations are a few percent and thus tolerable for a finite size scaling analysis.

With truncation, we can thus go further to larger systems via exact diagonalization. In Fig.~\ref{fig:scaling}, $n_i^{\mathrm{max}}=2$ is used for all the cases. Particle density $N/L$, the average number of particle per site, is fixed to $1/4, 1/3, 1/2, 1$ in the four panels of Fig.~\ref{fig:scaling}. For each filling factors, the trends of $\Delta_s$ for both strongly and weakly-interacting regimes are in agreement. The underlying understanding for the two typical trends is still unclear.  But we do know from Fig.~\ref{fig:scaling} that for systems of about 20 sites, the symmetry energy difference $\Delta_s$ is still remarkably large, at least for $N/L=1/3,1$. This indicates we should find critical symmetry breaking for larger systems of tens of sites.  We note that for larger system size, states with other rotational symmetry properties or with $m_n=+1$ rotational symmetry may enter the energy spectra where $\Delta_s$ extends, which could function as an internal noise source during the time evolution process.  Connections to open quantum systems for this case present an intriguing topic for future study.  The error bars in Fig.~\ref{fig:scaling}(c)(d) are obtained via the difference of $\Delta_s$ between $n_i^{\mathrm{max}}=3$ and $n_i^{\mathrm{max}}=2$ cases, as can of course only be done for smaller system sizes $L$ or particle number $N$ -- however, they do indicate the truncation error is relatively small, and our trends are indicative.  In fact, when the on-site interaction is strong (red curves), these differences are so slight that they can be disregarded, and error bars are smaller than the point size. In the weakly-interacting regime (blue curves), the errors are generally no more than 1\%.

\section{Conclusions and outlook}

In conclusion, we studied the nonequilibrium dynamics of bosons in a discrete optical ring trap or honeycomb lattice plaquette with an initial vortex state.  After quenching to a partial-symmetry broken lattice, we found critical behavior in the ensuing dynamics as determined by projection onto different rotational quantum numbers in the discretized system.  Up to a critical value of symmetry breaking, an initial winding number persists to long times; beyond this point, memory of the initial state is periodically lost and overall gradually decreases.  The \emph{symmetry memory}, or time average over such projections, was found to trend with the \emph{symmetry energy difference}, identified in the lowest lying cluster of excited states in the energy eigenspectrum.  Our exact diagonalization studies lay the groundwork for larger scale exploration of novel symmetry-based quantum dynamics in quantum simulator experiments.

A key question remains: how can our study of such small systems be extrapolated to the much larger system sizes present in quantum simulator experiments?  First, Fig.~\ref{fig:critlarge} indicates that larger system sizes also display symmetry breaking with the same two asymptotic trends in interaction strength as the 6-site case; thus according to the result from Fig.~\ref{fig:scaling} we expect that large ring optical lattices will lose memory of an initial vortex state for a critical symmetry breaking strength which remains sizable even for mesoscopic ring lattices of up to 20 sites.  Second, taking our 6-site case as a plaquette in a honeycomb lattice (see Fig.~\ref{fig:PSBtrap}) we can study the effect of vorticity distribution in a lattice system and how it depends on $A$-$B$ sublattice symmetry breaking; such biperiodic lattices were used previously for C-NOT gates in a square lattice~\cite{sebbeystrabley2006}. The results in this article with mesoscopic ring optical lattices of from 6 to 20 sites provide a foundation for future study of analogous processes in larger systems, wherever partial symmetry breaking occurs. Experiments with a honeycomb lattice may offer the advantage of performing the quantum average in a single shot since many plaquettes can display oscillation (or lack thereof) to different local winding number.  However, one should keep in mind that additional dynamics will appear due to coupling and overlap between plaquettes.  Connections to Berzinskii-Kosterlitz-Thouless (BKT) physics, where an ``infinite-order'' phase transition occurs due to locking of vortex-anti-vortex pairs, present an intriguing topic for future study, since partial symmetry breaking (in graphene terms, introducing an A-B sublattice gap) may block a BKT transition under the right circumstances.  In general, for the weakly interacting case and using the standard tool of interference with a condensate containing no vortices, such a distribution should appear as an array of bifurcations in the interference pattern~\cite{stock2006}.  Quantum microscopy may also prove useful for close observation~\cite{bakr2010} in order to determine winding number on hexagonal plaquettes in the honeycomb lattice~\cite{soltanPanahi2011}, in particular in the strongly interacting case where interference will not be sufficient due to lack of an emergent semiclassical condensate phase.  Due to the volume law associated with the quantum quench of partial symmetry breaking, this kind of study presents a totally new kind of quantum dynamics experiment requiring quantum supremacy~\cite{Boixo2016}, that is, inaccessible to quantum simulations on classical computers but perfectly accessible to quantum simulator experiments.

\ack{The authors thank Diego Alcala, Daniel Jaschke, Marc Valdez, and Biao Wu for helpful discussions. LDC acknowledges the University of Valencia for graciously hosting him on several occasions throughout work on this project. MAGM acknowledges the Fulbright commision, the Spanish Ministry MINECO (National Plan
15 Grant: FISICATEAMO No. FIS2016-79508-P, SEVERO OCHOA No. SEV-2015-0522), Fundaci\'o Privada Cellex, Generalitat de Catalunya (AGAUR Grant No. 2017 SGR 1341 and CERCA/Program), ERC AdG OSYRIS, EU FETPRO QUIC, and the National Science Centre, Poland-Symfonia Grant No. 2016/20/W/ST4/00314. This material is based in part upon work supported by Ministerio de Educación y Ciencia and EU FEDER under Contract FPA2016-77177 and by Generalitat Valenciana PrometeoII/2014/066 (JV), UV-INV-AE16-514545 and MINECO TEC2014-53727-C2-1-R (AF), the US National Science Foundation under grant numbers PHY-1806372, OAC-1740130, and CCF-1839232 (LDC), the US Air Force Office of Scientific Research grant number FA9550-14-1-0287 (LDC), the Alexander von Humboldt Foundation (LDC), the Heidelberg Center for Quantum Dynamics (LDC), and the China Scholarship Council (XXZ).

%\bibliography{PSB_refs_v5}
%\bibliographystyle{unsrt}

\end{document}